\documentclass[aps, twocolumn, groupedaddress,superscriptaddress, showpacs]{revtex4-1}
\usepackage{graphicx}
\usepackage{array}
\usepackage{setspace,transparent}
\usepackage{color}
\usepackage{fontenc,subcaption,ragged2e,amsmath}
\usepackage{mathtools,amssymb,amssymb,nicefrac,tikz}
\usepackage[font=small,labelfont=bf]{caption}
\usepackage[utf8]{inputenc}

\DeclareCaptionJustification{justified}{\justifying}
\newcommand{\RN}[1]{%
  \textup{\uppercase\expandafter{\romannumeral#1}}%
}
\DeclareMathAlphabet{\mathpzc}{OT1}{pzc}{m}{it}
\DeclareCaptionJustification{justified}{\justifying}
\begin{document}

%\title{External field effect on a first order phase transition for community structure under percolation process}
\title{Interconnections between networks act like an external field in first-order percolation transitions}
\author{Bnaya Gross}
\affiliation{Department of Physics, Bar Ilan University, Ramat Gan, Israel}
\author{Hillel Sanhedrai}
\affiliation{Department of Physics, Bar Ilan University, Ramat Gan, Israel}
\author{Louis Shekhtman}
\affiliation{Department of Physics, Bar Ilan University, Ramat Gan, Israel}
\author{Shlomo Havlin}
\affiliation{Department of Physics, Bar Ilan University, Ramat Gan, Israel}
\affiliation{Institute of Innovative Research, Tokyo Institute of Technology, Midori-ku, Yokohama, Japan 226-8503}
\date{\today}

\begin{abstract}
    %In many physical systems, aside from having coupling between the elements, there also exists some external field that effects all elements. The most well-known example of such a model is the  Ising model where an external magnetic field can be included. 
     Many interdependent, real-world infrastructures involve interconnections between different communities or cities. Here we study if and how the effects of such interconnections can be described as an external field for interdependent networks experiencing first-order percolation transitions. We find that the critical exponents $\gamma$ and $\delta$,  related to the external field can also be defined for first-order transitions but that they have different values than those found for second-order transitions. Surprisingly, we find that both sets of different exponents  can be found even within a single model of interdependent networks, depending on the dependency coupling strength. Specifically, the exponent $\gamma$ in the first-order regime (high coupling) does not obey the \textit{fluctuation dissipation} theorem, whereas in the continuous regime (for low coupling) it does. Nevertheless, in both cases they satisfy Widom's identity, $\delta - 1 = \gamma / \beta$ which further supports the validity of their definitions. {Our results provide physical intuition into the nature of the phase transition in interdependent networks and explain the underlying reasons for two distinct sets of exponents. }%Moreover, we find that in both interdependent networks and k-core percolation, the values of the critical exponents related to the field are the same suggesting that these systems belong to the same universality class.
    %These results elucidate the effects of an external field on first-order phase transitions and resilience properties of real-world networks.
\end{abstract}

\maketitle
In the past two decades, network theory has successfully described collective phenomena of many complex systems such as the brain, climate and infrastructures \cite{(7)-meunier2010modular,(7)-morone2017model, (7)-stam2010emergence, (7)-yamasaki2008climate}. The resilience of such networks is often studied under a percolation process where a fraction $1 - p$ of nodes are removed randomly from the network and the size of the largest connected component, $S$ (the order parameter of the system) is measured \cite{bunde2012fractals,albert2000error,cohen-prl2000,cohenhavlin2010complex_book,newman2011structure_book,newmancallaway2000network}. %It has been found that for many networks there exists a phase transition in $S$ as $p$ is reduced from $1$ to some critical value, $p_c$. Below this transition point, the size of the giant component is essentially zero. The value of $p_c$ and the critical exponents characterizing the nature of the transition are also known to depend on the structure of the network including features like the degree distribution of nodes \cite{cohen2002percolation}, possible spatial embedding \cite{daqing_2011}, and other properties. 
Several generalizations of percolation-like processes have also been developed and these processes also affect the values of the critical exponents, and the nature of the transition e.g., whether it is continuous or abrupt. 
%{\color{red} Specifically, $k$-core percolation, is an iterative process where nodes with less than $k$ neighbors are considered failed such that in the final giant component of all remaining nodes have at least $k$ links to other surviving nodes \cite{dorogovtsev2006k,goltsev2006k,carmi2007model,watts2002simple_pnas}. 
Specifically, \emph{interdependent} networks \cite{buldyrev-nature2010,gao2012interdependentnetworks,rinaldi-ieee2001,gao2011robustnessNoN} where one network depends on another, have drawn much interest. In these systems there exist several networks with the ordinary connectivity links within the networks, yet dependency links between the networks imply that if a node at one end of a dependency link fails than the node at the other end will also fail, even if it is still connected in its own network. This process leads percolation on interdependent networks to result in cascading failures leading to abrupt, first-order percolation transitions. 

Several researchers have studied  `interconnected networks' where two networks each with many connections inside their own network, also have a smaller number of links between them \cite{saumell2012epidemic,radicchi-naturephysics2013,de2014navigability,brummitt2012suppressing_raissa_dsouza_pnas}. Such networks have also been described in the literature as networks with community structure, since each network can be regarded as a separate community \cite{girvan2002community,palla2005uncovering,lancichinetti2009detecting,mucha2010community}. %\textcolor{red}{(BG: I keep this paragraph here till we agree what to do with it) LMS: I like this paragraph and think it is important to also mention some of the prior work on communities. Also in any case we should cite these people}. 
\begin{figure}
	\centering
	\begin{tikzpicture}[      
	every node/.style={anchor=north east,inner sep=0pt},
	x=1mm, y=1mm,
	]   
	\node (fig1) at (-30,0)
	{\includegraphics[scale=1]{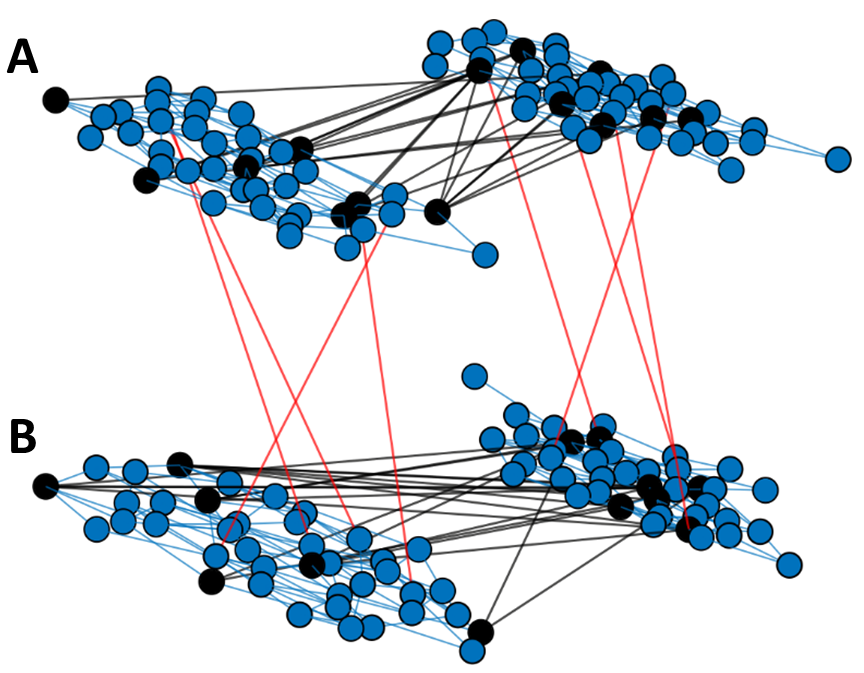}};
	\end{tikzpicture}
	\caption{\textbf{Model illustration.} The model is composed of two networks A and B, each one contain two modules (\textbf{\textcolor{blue}{blue}} nodes) with a small fraction $r$ of nodes that have interlinks connecting the two modules (\textbf{black} nodes). $M_{inter}$ links are assigned randomly between the cores of the modules (\textbf{black} links). The two networks A and B depend on each other via a fraction $q$ of dependency links shown in \textbf{\textcolor{red}{red}}. The dependency links are restricted to be within the same community in each network.} % Each network has $N$ nodes and the system has $4N$ nodes total.}
	\label{fig:Illustration}	
\end{figure}
Recently, a new model of community structure has been proposed where only some small fraction, $r$, of nodes are assumed \emph{a priori} to be capable of having interlinks to other communities \cite{Gaogaodong2018resilience}. It was found that the fraction $r$ of such nodes affects the percolation transition in a manner analogous to an external field in spin systems or a ghost field in percolation. \par

The effects of an external field are best characterized through the key exponents $\beta$, $\delta$, and $\gamma$ describing the behavior of the system near (and at) criticality \cite{bunde2012fractals,(1-6)-huang2009introduction,(1-4-6)-stanley1971phase,(4-5)-stauffer2014introduction}, which fulfill Widom's identity $\delta - 1 = \gamma / \beta$ implying that there are only 2 degrees of freedom in determining these exponents. For percolation processes, these critical exponents are defined when the control parameter $p$ is near (and at) the percolation threshold $p_c$. \\
$\mathit{1)}$ The critical exponent $\beta$ describes the behavior of the order parameter ($S$) near the critical point with zero-field ($r=0$) and is given by
\begin{equation}
    S(0,p) - S(0,p_c) \sim (p - p_c)^{\beta} \quad .
\end{equation}
$\mathit{2)}$ \textit{At the critical point,} ($p=p_c$), the increase of the order parameter with the magnitude of the field, $r$, is given by the critical exponent $\delta$ as
\begin{equation}
    S(r,p_c) - S(0,p_c) \sim r^{1/\delta} \quad .
\end{equation}
$\mathit{3)}$ The susceptibility of the system, $\chi$, is given by the partial derivative of the order parameter with respect to the field, $r$, and scales near the critical point with the exponent $\gamma$ as 
\begin{equation}
    \chi \equiv \left(\frac{\partial S(r,p)}{\partial r}\right)_{r \rightarrow 0} \sim |p - p_c|^{-\gamma} \quad  .
\end{equation}

Here we study analytically and via simulations the percolation of this novel community structure in interdependent networks (see Fig. \ref{fig:Illustration}) with $q$ fraction of interdependent nodes i.e.,  $1-q$ fraction of nodes in each network are autonomous. We observe two distinct regimes in this single model, characterizing different values of the critical exponents and universality class.  For small values of $q$, the network undergoes a continuous second-order phase transition as for isolated ER networks and has the corresponding critical exponent values.
However, for large values of $q$ these systems undergo first-order phase transitions with a different set of exponent values. Moreover the fraction of interconnected nodes, $r$, can even for the case of an abrupt transition, be analogized to an external field. Indeed, prior work gave puzzling results in the first-order transition regime, where two alternative definitions of the exponent $\gamma$ were shown to lead to two distinct exponent values \cite{lee2016hybrid}.  Here, by using our new definition of relating interconnections to an external field, we demonstrate that this puzzle can be rectified by recognizing that these distinct values imply a violation of the fluctuation dissipation theorem.

%We study a model of \textit{interdependent networks} \cite{parshani-prl2010} each with the community structure described above (see Fig.~\ref{fig:Illustration}) and also explore the limiting case of $q \to 1$ representing fully interdependent networks, in which $\gamma$ was first shown to have different values for different definitions. %We also study the $k$-core percolation on a \textit{single} interconnected network (top or bottom of  Fig. \ref{fig:Illustration}).  
\par
% Even though the effect of an external field have been studied extensively over the years, the effect was only examined on systems with second order phase transitions. In this paper, we would study the effect of an external field on \textit{first order} phase transition using the external field recently found for community structure under percolation process \cite{Gaogaodong2018resilience}.
% {\color{red} LMS: I am unconvinced on this introduction and title. I think we should discuss with Shlomo when he comes back whether to make the main point be `finding scaling for a first-order phase transition' or make it be more from a network perspective with the first-order scaling being an additional nice point, but not necessarily highlighting the connection to very old literature.}
\underline{\textit{Model.--}} Our network model assumes two communities where only a small fraction $r$ of nodes in each community are capable of having interlinks \cite{Gaogaodong2018resilience}. A total of $M_{inter}$ links are then assigned among this small subset of nodes. Motivation for the interconnected structure of the model can be found in systems where additional resources are needed at a site in order to accommodate long-range links. At the same time, once such infrastructure exists, adding additional interlinks is of low cost. One example is the airport network where longer runways are needed for planes that fly transoceanic flights and thus some airports have such flights while others do not. Nonetheless, once this infrastructure exists, adding more transoceanic flights is easy.  Similarly, power stations that transfer large load to long distances may require additional infrastructure in order to handle such load. Next, two networks constructed from this model are set to be partially interdependent with $q$ fraction of nodes in each network depending on nodes in the other network, as seen in Fig. \ref{fig:Illustration}. 

\underline{\textit{Analytic solution.--}} %We know that for $q=0$, the system must have the same scaling exponents as a single ER network \cite{Gaogaodong2018resilience} while for $q=1$ the exponents are different \cite{parshani-prl2010}. Thus, we now choose to study how, in a single model incorporating $q$, these two sets of exponents arise. \\
We begin by developing an analytic solution for the effect of interlinks on percolation of  interdependent networks of the type described above. We start by defining the generating functions for the degree distribution of intra- and inter- connected nodes. For intra nodes we obtain $G^{intra}_0(x) = \sum_{k} p^{intra}_kx^k$ and $G^{intra}_1(x) = \sum_{k} q^{intra}_kx^k$ where $p^{intra}_k$ is the probability for a node to have $k$ intra links and $q^{intra}_k = \frac{(k+1)p^{intra}_{k+1}}{z}$ is the intra excess degree distribution with $z$ being the average intra degree \cite{newmancallaway2000network,newman2001random}.  We assume that interlinks are always assigned randomly and thus their generating functions are given by $G^{inter}_0(x) = G^{inter}_1(x) = e^{-\kappa(1-x)}$ where $\kappa = \frac{M_{inter}}{rN}$ is the average inter-degree of the $r$ fraction of nodes in the core and $N$ is the total number of nodes. \par
\begin{figure}
	\centering
	\begin{tikzpicture}[      
	every node/.style={anchor=north east,inner sep=0pt},
	x=1mm, y=1mm,
	]   
	\node (fig1) at (0,0)
	{\includegraphics[scale=0.37]{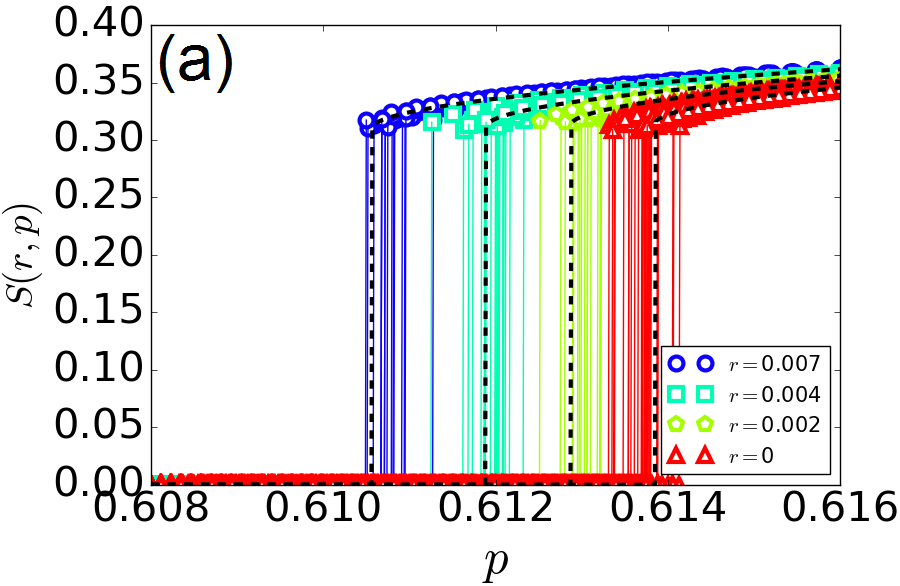}};
	\node (fig1) at (-53,-22)
	{\includegraphics[scale=0.17]{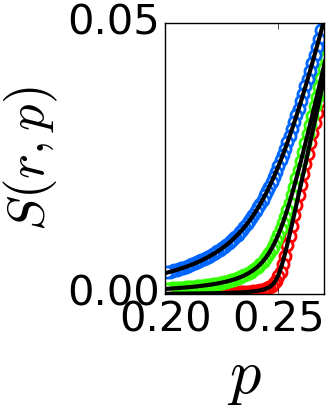}};
	\node (fig1) at (-53,-60)
	{\includegraphics[scale=0.19]{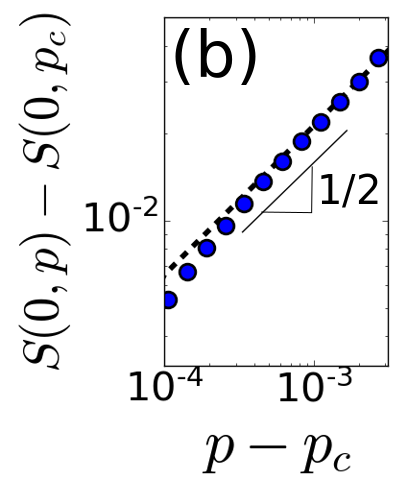}};
	\node (fig1) at (-24,-60.3)
	{\includegraphics[scale=0.19]{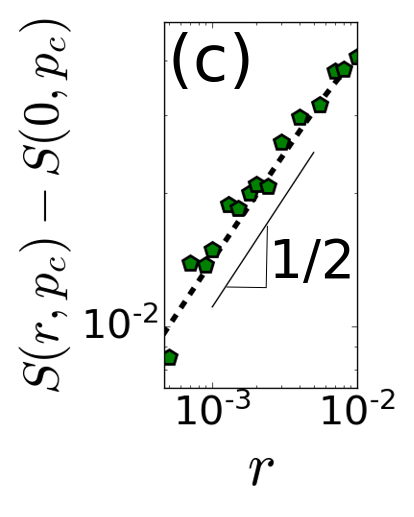}};
	\node (fig1) at (3.2,-60.5)
	{\includegraphics[scale=0.16]{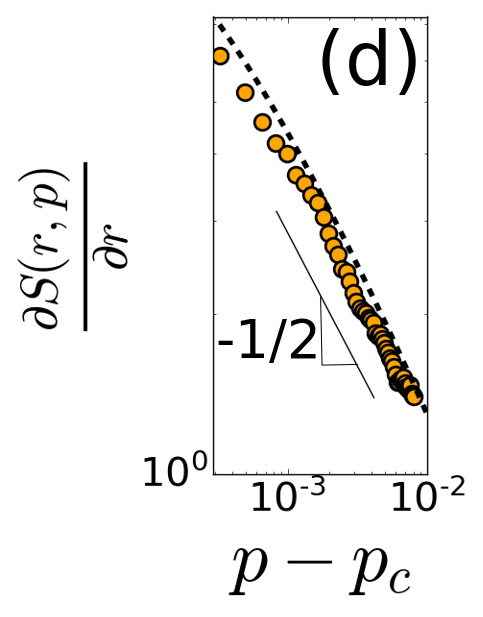}};
	\end{tikzpicture}
	\caption{\textbf{(a) The size of the mutual giant connected component $S(r,p)$ for interdependent networks with large $q$ whose communities each have an ER structure.} Simulations of single realizations (colored symbols) and theory (dashed lines) from Eq. \eqref{eq:theory_giant_interdependent}, are shown. At %for the size of the giant component $S(r,p)$. 
	the  critical point of the system with no field, $p_c(r=0)$, we observe that as $r$ increases, $S(r,p_c)$ increases as well. The power-law scaling of this increase suggests that $r$ can be analogized to an external field. The inset shows the effect of the external field, $r$, for a single network composed of two communities and a continuous transition (not interdependent). Simulations are shown for $N=10^7$ nodes, mean degree $z = 4$ and $M_{inter}=N/20$. We find that \textbf{The critical exponents} have the values \textbf{(b)} $\beta = 1/2$, \textbf{(c)} $\delta = 2$ and \textbf{(d)} $\gamma = 1/2$ and the Widom's identity $\delta - 1 = \gamma/ \beta$ is satisfied. The simulations are plotted with symbols and the theory as dashed lines. The simulations for $\gamma$ are shown for $N=10^8$ and $r = 0.0007$.}
	\label{fig:ER}	
\end{figure}

We next define $u$ and $v$, the probability that after removal of $1-p$ fraction of nodes from \textit{each} network, an intra and inter edge respectively do not lead to a node connected to the giant component. They satisfy the equations:
\begin{gather*}
	u =1-p\left[1-G^{intra}_1(u)(1-r+rG^{inter}_0(v))\right] \times \\ \times \left[1-q+qp(1 - G^{intra}_0(u)(1-r+rG^{inter}_0(v))\right], \\
	v = 1-p\left[1-G^{intra}_0(u)G^{inter}_1(v)\right] \times \\
	\times \left[ 1-q + qp(1-G^{intra}_0(u)G^{inter}_0(v))\right].
	\label{eq:uv_q}
\end{gather*}
For ER networks (i.e., $p_k^{intra} = \frac{z^ke^{-z}}{k!}$), $G^{intra}_0(u) = G^{intra}_1(u) = e^{-z(1-u)}$ and $S=1-u$, leading to a single transcendental equation relating $S$, $q$, and $r$
\begin{widetext}
\begin{equation}
	\frac{1-q+2qp - \sqrt{(1-q)^2+4qS}}{2qp}e^{zS}+(r-1) 
	= r\exp\Bigg(\frac{\kappa p}{r}\bigg[(r-1)(e^{-zS}-1) \frac{1-q+\sqrt{(1-q)^2+4qS}}{2} -\frac{S}{p}\bigg] \Bigg).
	\label{eq:S_q}
\end{equation}
\end{widetext}
As seen in Fig. \ref{fig:ER}, for large values of $q$ the system undergoes an abrupt first order transition while for small values of $q$ (Fig. \ref{fig:ER} inset and \cite{Gaogaodong2018resilience,lee2016hybrid}) it experiences a continuous second-order transition. As $r$ increases the size of the mutual giant component at $p_c(r=0)$ increases with clear scaling relationships (see below and Fig. \ref{fig:ER}b,c,d) suggesting that $r$ can be analogized to an external field. Two sets of different critical exponents arise from Eq. \eqref{eq:S_q}. For strong dependency (i.e. large values of $q$) we obtain $\delta = 2$ and $\beta = \gamma = 1/2$ while for weak coupling we find $\delta = 2$ and $\beta = \gamma = 1$ \cite{Gaogaodong2018resilience}. Both sets of critical exponents satisfy Widom's identity $\delta - 1 = \gamma / \beta$.
\par
For strong dependency, our results also support prior results by \textit{Lee. et al} for the value of $\beta$ \cite{lee2016hybrid}. However, our results differ from the results of \cite{lee2016hybrid} who obtained $\gamma=1$ when using a definition of $\gamma$ based on the fluctuations in the size of the order parameter ($S$) near criticality indicating a \textit{violation of the fluctuation dissipation theorem} (FDT). We resolve this contradiction by suggesting that the reason for this phenomena is fundamental since the FDT holds only in equilibrium while for systems out of equilibrium (e.g., undergoing an abrupt transition  due to the spreading of damage) as here, there are no guarantees of its validity. Indeed, violations of the FDT have been also observed and studied widely in glassy systems \cite{crisanti2003violation_FDT,grigera1999observation_VFDT,marinari1998violation_FDT} due to the aging phenomena (remaining out of equilibrium for all time). Indeed, our system exhibits a first order transition with a metastable state which while often treated as if it is an equilibrium state, it is not. Thus our result support the suggestion in \cite{baez2003fluctuation_FDT_metastable} that first order transitions are off equilibrium. This therefore leads to a violation of the FDT. 
\par
\begin{figure}
	\centering
	\begin{tikzpicture}[      
	every node/.style={anchor=north east,inner sep=0pt},
	x=1mm, y=1mm,
	]   
	\node (fig1) at (0,0)
	{\includegraphics[scale=0.25]{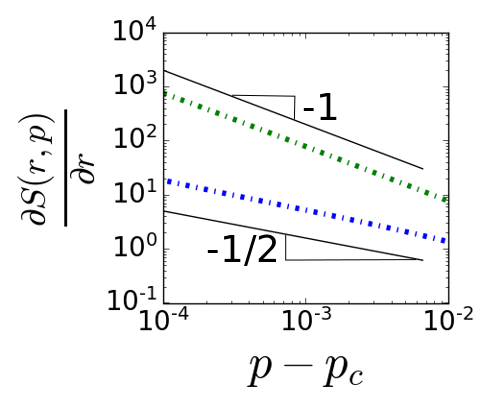}};
	\node (fig1) at (-45,-0.5)
	{\includegraphics[scale=0.25]{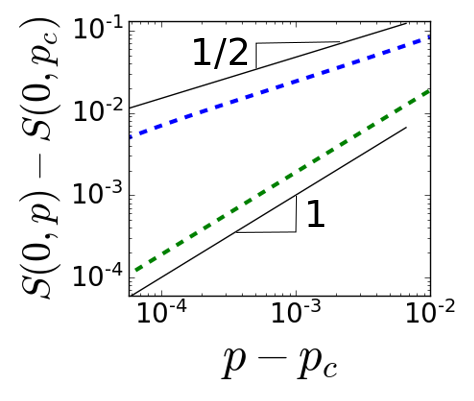}};
	\end{tikzpicture}
	\caption{\textbf{The critical exponents $\beta$ (left) and $\gamma$ (right) for $q = 0.2$ (green) and $0.8$ (blue)}. It can be seen that for small values of $q$ we get $\beta = \gamma = 1$, while for large values of $q$ we get $\beta = \gamma = 1/2$. Here $r = 10^{-8}$ .}
	\label{fig: ER_q_gamma_beta}	
\end{figure}

%\begin{multline}
% 	\frac{1-q+2qp - \sqrt{(1-q)^2+4qS}}{2qp}e^{zS}+(r-1) \\
% 	= r\exp\Bigg(\frac{\kappa p}{r}\bigg[(r-1)(e^{-zS}-1)\times \\ \frac{1-q+\sqrt{(1-q)^2+4qS}}{2} -\frac{S}{p}\bigg] \Bigg).
% 	\label{eq:S_q}
% \end{multline}
% LMS: If you dont like the widetext then comment it out and use the multline.
% Indeed, note that for $q=1$, Eq. \eqref{eq:theory_giant_interdependent} is recovered (with $p \rightarrow \sqrt{p}$ since there we only removed $p$ fraction of nodes from a single network), while for $q = 0$ the equation for a single network is recovered \cite{Gaogaodong2018resilience}.
In Fig. \ref{fig: ER_q_gamma_beta}, the two sets of different  critical exponents for strong and weak dependency are obtained from numerical analysis of Eq. \eqref{eq:S_q}. Later we present also the analytic derivation of the two sets of exponents based on Eq. \eqref{eq:S_q}. It can be seen that for large values of $q$ we obtain $\gamma = 1/2$, the FDT is violated and the system undergoes an abrupt transition (Fig. \ref{fig:ER}). In contrast, for small values of $q$ we obtain $\gamma = 1$ and the FDT holds with the system undergoing a continuous second-order transition (Fig. \ref{fig:ER} inset and \cite{Gaogaodong2018resilience,lee2016hybrid}). This shows how even in a single model (represented by a single equation (Eq. \eqref{eq:S_q})) we can observe both sets of exponents and further supports our claim that the violation of the FDT is inherent to this type of transition and not related to differences between models. \par
\underline{\textit{Limit of $q = 1$} (Fully interdependent Networks).--} For the specific case of fully interdependent networks ($q = 1$), we consider the case of removal of $1 - p$ fraction of nodes from \textit{network A only} in order to be consistent with \cite{buldyrev2010catastrophic,gao2012interdependentnetworks}. In this case $u$ and $v$ satisfy the equations \cite{Gaogaodong2018resilience}:
\begin{gather*}
	u =1-p+pG^{intra}_1(u)[1-r+rG^{inter}_0(v)],\\
	v = 1-p+pG^{intra}_0(u)G^{inter}_1(v) .
	\label{eq:uv}
\end{gather*}
%Likewise, $u$ and $v$ can be used to obtained the size of the giant component for a single isolated network \cite{Gaogaodong2018resilience}.
% The size of the giant component for a single isolated network is then \cite{Gaogaodong2018resilience}:
% \begin{equation}
% 	S = pg(p) = p(1 - G^{intra}_0(u)[1-r+rG^{inter}_0(v)]) .
% 	\label{eq:giant_single}
% \end{equation}
% For ER networks (i.e., $p_k^{intra} = \frac{z^ke^{-z}}{k!}$), $G^{intra}_0(u) = G^{intra}_1(u) = e^{-z(1-u)}$, which together with the above definitions of $u$ and $v$ can be used to obtain the size of the giant component for a single network. Simulations and theory for a single network show excellent agreement in the inset of Fig. \ref{fig:ER}, confirming the prior work of Dong et al. \cite{Gaogaodong2018resilience}. Note that for $r=0$ the size of the giant component at criticality is zero and for $r > 0$ it becomes non-zero. %{\color{red} LMS: I suggest deleting these two above sentences and perhaps even the equation. This is just repeating Gaogao's work.}
\par
\begin{figure}
	\centering
	\begin{tikzpicture}[      
	every node/.style={anchor=north east,inner sep=0pt},
	x=1mm, y=1mm,
	]   
	\node (fig1) at (0,-80)
	{\includegraphics[scale=0.31]{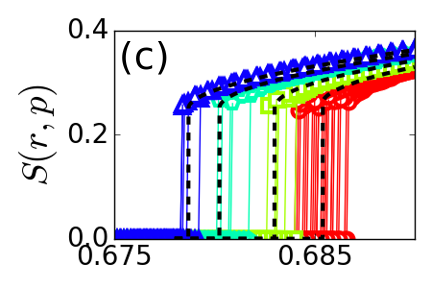}};
	\node (fig1) at (0,-40)
	{\includegraphics[scale=0.30]{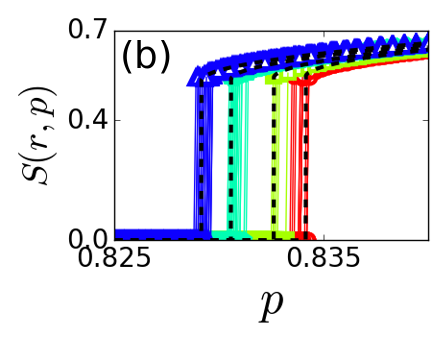}};
	\node (fig1) at (0,0)
	{\includegraphics[scale=0.32]{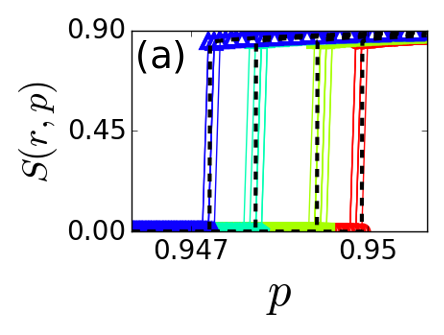}};
	\node (fig1) at (42,-82)
	{\includegraphics[scale=0.23]{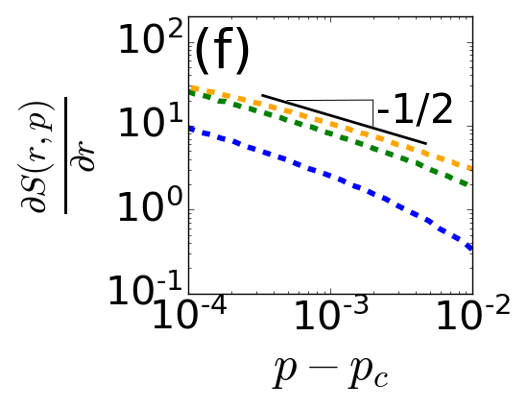}};
	\node (fig1) at (42.5,-40.5)
	{\includegraphics[scale=0.22]{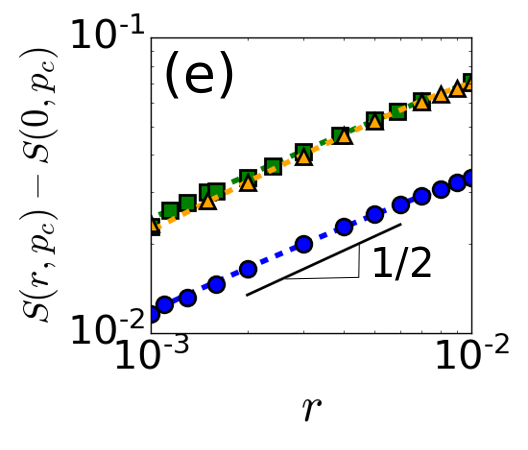}};
	\node (fig1) at (42,-2.5)
	{\includegraphics[scale=0.18]{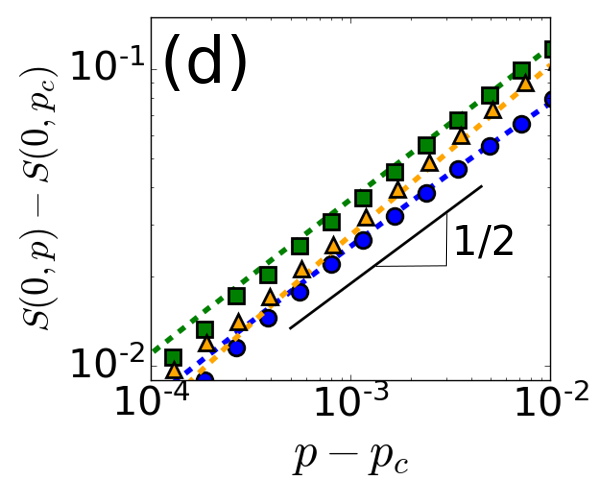}};
	\end{tikzpicture}
	\caption{\textbf{The size of the mutual giant connected component $S(r,p)$ for interdependent SF networks with large $q$.} \textbf{(a)} $\lambda = 4.5$, \textbf{(b)} $\lambda = 3.35$ and \textbf{(c)} $\lambda = 2.8$. Here $r = 0.01$ (blue triangles), $r = 0.007$ (cyan pentagons), $r = 0.003$ (yellow squares) and $r = 0$ (red circles). \textbf{The critical exponents} \textbf{(d)} $\beta = 1/2$, \textbf{(e)} $\delta = 2$ and  \textbf{(f)} $\gamma = 1/2$. Colors and symbols are $\lambda = 4.5$ - blue circles,  $\lambda = 3.35$ - green squares and $\lambda = 2.8$ - orange triangles. In all cases $\beta$, $\delta$ and $\gamma$ are the same as for ER (see Fig. \ref{fig:ER}) suggesting that for interdependent networks ER and SF are in the same universality. Simulations are shown for $N=10^7$ and $M_{inter} = N/20$. The theory in all the plots is shown as dashed lines. For $\gamma$ the theory is shown for $r = 0.0001$.}
	\label{fig:SF}	
\end{figure}

For the case of fully interdependent ER networks we use the framework from \cite{buldyrev2010catastrophic,gao2012interdependentnetworks} to arrive at the following equation for the mutual giant component (see SI), %{\color{red} LMS: I removed the two lines about self-consistency equations as it is overly technical and not needed. Please check that this derivation is actually covered in the SI.}
\begin{multline}
e^{-zS}(r-1) + 1 - \sqrt{\frac{S}{p}} = \\
= r\exp\left[\frac{\kappa \sqrt{Sp}(e^{-zS}(r-1) + 1 - \sqrt{\frac{S}{p}} -r)}{r} -zS\right] .
\label{eq:theory_giant_interdependent}
\end{multline}
Note that for $r=0$, Eq. \eqref{eq:theory_giant_interdependent} recovers the well-known result for two interdependent networks $S = p(1-e^{-zS})^2$ and Eq. \eqref{eq:S_q} is recovered with $p \rightarrow \sqrt{p}$ since there we removed $p$ fraction of nodes from both networks. Fig. \ref{fig:ER} shows excellent agreement between the theory of Eq. \eqref{eq:theory_giant_interdependent} and simulations and the set of critical exponents for strong dependency is presented in Fig. \ref{fig:ER}b,c,d. Fig. \ref{fig:SF} shows $S(r,p)$ for interdependent SF networks (i.e., $p_k^{intra} \sim k^{-\lambda}$) for different values of $\lambda$ with large $q$ and shows perfect agreement between the theory and the simulations. We also observe here a clear scaling relationship between $r$ and $S$ (Fig. \ref{fig:SF}d,e,f), further justifying the analogy to an external field.\par
%  The theory is evaluated from Eqs. \eqref{eq:uv} using the equations for fully interdependent networks and agrees perfectly with the simulations
\underline{\textit{Analytic Derivation of Critical Exponents.--}} Having solved the model and demonstrated that scaling relationships analogous to those of an external field can be found, here we seek to extract these scaling exponents analytically. This can be done by assuming $\kappa$ to be a constant. To derive the exponents for ER networks, we let $f(S,p,r,q)$ be
\begin{multline}
   f(S,p,r,q) = \bigg[(1-r)(1 - e^{-zS}) \cdot \\
   \cdot \frac{1 - q + \sqrt{(1 - q)^2 + 4qS}}{2} - \frac{S}{p} \bigg],
\end{multline}
and thus Eq. \eqref{eq:S_q} takes the form
\begin{multline}
   r + \frac{2f(S,p,r,q)}{1-q+\sqrt{(1-q)^2 + 4qS}} = \\
   = r\exp \bigg[\frac{\kappa p f(S,p,r,q)}{r} - zS\bigg].
\end{multline}
In the limit of $r\to 0$, we recover $f(S,p,0,q) = 0$ defining the giant component of two interdependent ER networks.
Likewise, at criticality, $f_{S}(S_c,p_c,0,q) = 0$, where $f_S$ refers to the derivative of $f$ with respect to $S$. Expanding $f(S,p,0,q)$ around $S=S_c$ and $p=p_c$ gives
\begin{equation*}
    \begin{split}
        & f(S,p,0,q) = f(S_c,p_c,0,q)+f_S(S_c,p_c,0,q)(S - S_c) +  \\
        & + f_p(S_c,p_c,0,q)(p-p_c) + ... = 0 .
    \end{split}
\end{equation*}
Further developing this expansion leads to $\beta = 1/2$ for large $q$ and $\beta = 1$ for small $q$ analytically. The exponents $\delta$ and $\gamma$ require different approximations in the expansion of $f(S,p,r,q)$ in order to arrive at their values analytically (we find $\delta = 2$ and $\gamma = 1/2$ for large $q$ and $\delta = 2$ and $\gamma = 1$ for small $q$), see SI for the detailed derivation. 
Simulations and theory for the critical exponents for ER networks with large $q$ are in excellent agreement and are shown in Fig. \ref{fig:ER}b-d. The simulations support our derivations that $\delta = 2$ and $\beta = \gamma=1/2$. These values also satisfy Widom's identity. %and are identical to those found for $k$-core percolation \cite{dorogovtsev2006k,goltsev2006k,carmi2007model} on a Bethe lattice \cite{schwarz2006onset_kcore_field}. 
We also confirmed these values for $k$-core percolation \cite{dorogovtsev2006k,goltsev2006k,carmi2007model,schwarz2006onset_kcore_field} on networks with the community structure defined here (see SI) and found the same values--  further supporting that interdependent networks and $k$-core percolation are in the same universality class. Simulations for the critical exponents for small $q$ are shown else where \cite{Gaogaodong2018resilience} and also show excellent agreement with our derivation that $\delta = 2$ and $\beta = \gamma = 1$. Thses values also satesfy Widom's identity.\par
%which already been shown for glassy systems \cite{crisanti2003violation_FDT,grigera1999observation_VFDT,marinari1998violation_FDT} and domain growth systems \cite{barrat1998monte_VFDT}. In the discussion section we will address broadly to this phenomenon.
We also measure the critical exponents of the external field for SF networks having different values of $\lambda$ for large $q$. The results for $S(r,p)$ and the critical exponents for different values of $\lambda$ are shown in Fig. \ref{fig:SF}. For $\lambda > 4$ we have low heterogeneity and indeed as expected we find similar results as for ER networks (i.e, $\delta = 2$ and $\beta=\gamma = 1/2$). For high heterogeneity (i.e. $3<\lambda<4$ and $2<\lambda<3$) the system has different exponents than ER for the case of single layers \cite{Gaogaodong2018resilience}, yet for interdependent networks we find that the critical exponents are the same as for interdependent ER networks, $\delta = 2$ and $\beta = \gamma = 1/2$. Thus, our results suggest that interdependent ER networks and interdependent SF networks with large $q$ are in the same universality class in contrast to small $q$ at which the exponents are different \cite{Gaogaodong2018resilience}. The reasoning is  most probably due to the fact that the random spread of damage due to interdependence does not distinguish between high and low degree nodes. %Note that the same is true for k-core percolation as discussed below.
\par
In summary, we have studied the effects of an external field on first-order percolation phase  transitions by analyzing analytically and numerically, interdependent networks with interconnections. We find that a model of interdependent networks is able of expressing two characteristic behaviors depending on the level of interdependence coupling, $q$. For high-values of $q$ the critical exponents are the same for both ER and SF networks ($\delta = 2$ and $\beta = \gamma = 1/2$). These exponents satisfy Widom's identity $\delta - 1 = \gamma / \beta$ and their common value suggests the existence of a single universality class describing these cascading phenomenon. Interestingly, we observe a \textit{violation of the fluctuation dissipation theorem} (FDT) \cite{kubo1966fluctuation_FDT,weber1956fluctuation_FDT}, expressed through the contradiction between the values of the critical exponent $\gamma$ for different descriptions of the susceptibility $\chi$. This violation helps solving the puzzle of why a field response description yields $\gamma = 1/2$ while a fluctuation description gives $\gamma = 1$ \cite{lee2016hybrid}. Our model displays both behaviors, as for strong coupling the FDT is violated, while for weaker coupling it holds, further demonstrating that this is the previously unknown reason for the different exponent values.

% The origin of this phenomena inherent is since FDT holds only in equilibrium while for systems out of equilibrium (abrupt transition  due to spreading of damage) there is no guarantee for it to hold. Indeed, violations of the FDT were observed and studied widely in glassy systems \cite{crisanti2003violation_FDT,grigera1999observation_VFDT,marinari1998violation_FDT} due to the aging phenomena (staying out of equilibrium for all times). Our system exhibits a first order transition, a metastable state which is usually treated as if it is an equilibrium state but in fact it is not as suggested in \cite{baez2003fluctuation_FDT_metastable} causing violation of the FDT to occur. \par
\underline{\textit{Acknowledgments.--}} B. G. and H. S. contributed equally to this work. We thank Ivan Bonamassa for very useful discussions related to this project. We also thank the Italian Ministry of Foreign Affairs and International Cooperation jointly with the Israeli Ministry of Science, Technology, and Space (MOST); the Israel Science Foundation, ONR, the Japan Science Foundation with MOST, BSF-NSF, ARO, the BIU Center for Research in Applied Cryptography and Cyber Security, and DTRA (Grant no. HDTRA-1-10-1-0014) for financial support.
\bibliographystyle{unsrt}
\bibliography{NoN}

\begin{thebibliography}{10}

\bibitem{(7)-meunier2010modular}
David Meunier, Renaud Lambiotte, and Edward~T Bullmore.
\newblock Modular and hierarchically modular organization of brain networks.
\newblock {\em Frontiers in neuroscience}, 4:200, 2010.

\bibitem{(7)-morone2017model}
Flaviano Morone, Kevin Roth, Byungjoon Min, H~Eugene Stanley, and Hern{\'a}n~A
  Makse.
\newblock Model of brain activation predicts the neural collective influence
  map of the brain.
\newblock {\em Proceedings of the National Academy of Sciences},
  114(15):3849--3854, 2017.

\bibitem{(7)-stam2010emergence}
Cornelis~Jan Stam, Arjan Hillebrand, Huijuan Wang, and Piet Van~Mieghem.
\newblock Emergence of modular structure in a large-scale brain network with
  interactions between dynamics and connectivity.
\newblock {\em Frontiers in computational neuroscience}, 4:133, 2010.

\bibitem{(7)-yamasaki2008climate}
Kazuko Yamasaki, Avi Gozolchiani, and Shlomo Havlin.
\newblock Climate networks around the globe are significantly affected by el
  nino.
\newblock {\em Physical Review Letters}, 100(22):228501, 2008.

\bibitem{bunde2012fractals}
Armin Bunde and Shlomo Havlin.
\newblock {\em Fractals and disordered systems}.
\newblock Springer Science \& Business Media, 2012.

\bibitem{albert2000error}
R{\'e}ka Albert, Hawoong Jeong, and Albert-L{\'a}szl{\'o} Barab{\'a}si.
\newblock Error and attack tolerance of complex networks.
\newblock {\em Nature}, 406(6794):378--382, 2000.

\bibitem{cohen-prl2000}
Reuven Cohen et~al.
\newblock {Resilience of the Internet to Random Breakdowns}.
\newblock {\em Phys. Rev. Lett.}, 85:4626--4628, Nov 2000.

\bibitem{cohenhavlin2010complex_book}
Reuven Cohen and Shlomo Havlin.
\newblock {\em Complex networks: structure, robustness and function}.
\newblock Cambridge university press, 2010.

\bibitem{newman2011structure_book}
Mark Newman, Albert-Laszlo Barabasi, and Duncan~J Watts.
\newblock {\em The structure and dynamics of networks}, volume~19.
\newblock Princeton University Press, 2011.

\bibitem{newmancallaway2000network}
Duncan~S Callaway, Mark~EJ Newman, Steven~H Strogatz, and Duncan~J Watts.
\newblock Network robustness and fragility: Percolation on random graphs.
\newblock {\em Physical Review Letters}, 85(25):5468, 2000.

\bibitem{buldyrev-nature2010}
Sergey~V. Buldyrev et~al.
\newblock {Catastrophic cascade of failures in interdependent networks}.
\newblock {\em Nature}, 464(7291):1025--1028, Apr 2010.

\bibitem{gao2012interdependentnetworks}
Jianxi Gao, Sergey~V Buldyrev, H~Eugene Stanley, and Shlomo Havlin.
\newblock Networks formed from interdependent networks.
\newblock {\em Nature Physics}, 8(1):40, 2012.

\bibitem{rinaldi-ieee2001}
S.M. Rinaldi, J.P. Peerenboom, and T.K. Kelly.
\newblock {Identifying, understanding, and analyzing critical infrastructure
  interdependencies}.
\newblock {\em Control Systems, IEEE}, 21(6):11--25, 2001.

\bibitem{gao2011robustnessNoN}
Jianxi Gao, Sergey~V Buldyrev, Shlomo Havlin, and H~Eugene Stanley.
\newblock Robustness of a network of networks.
\newblock {\em Physical Review Letters}, 107(19):195701, 2011.

\bibitem{saumell2012epidemic}
Anna Saumell-Mendiola, M~{\'A}ngeles Serrano, and Mari{\'a}n Bogun{\'a}.
\newblock Epidemic spreading on interconnected networks.
\newblock {\em Physical Review E}, 86(2):026106, 2012.

\bibitem{radicchi-naturephysics2013}
Filippo Radicchi and Alex Arenas.
\newblock {Abrupt transition in the structural formation of interconnected
  networks}.
\newblock {\em Nature Physics}, 9(11):717--720, Sep 2013.

\bibitem{de2014navigability}
Manlio De~Domenico, Albert Sol{\'e}-Ribalta, Sergio G{\'o}mez, and Alex Arenas.
\newblock Navigability of interconnected networks under random failures.
\newblock {\em Proceedings of the National Academy of Sciences},
  111(23):8351--8356, 2014.

\bibitem{brummitt2012suppressing_raissa_dsouza_pnas}
Charles~D Brummitt, Raissa~M D’Souza, and Elizabeth~A Leicht.
\newblock Suppressing cascades of load in interdependent networks.
\newblock {\em Proceedings of the National Academy of Sciences},
  109(12):E680--E689, 2012.

\bibitem{girvan2002community}
Michelle Girvan and Mark~EJ Newman.
\newblock Community structure in social and biological networks.
\newblock {\em Proceedings of the National Academy of Sciences},
  99(12):7821--7826, 2002.

\bibitem{palla2005uncovering}
Gergely Palla, Imre Der{\'e}nyi, Ill{\'e}s Farkas, and Tam{\'a}s Vicsek.
\newblock Uncovering the overlapping community structure of complex networks in
  nature and society.
\newblock {\em Nature}, 435(7043):814--818, 2005.

\bibitem{lancichinetti2009detecting}
Andrea Lancichinetti, Santo Fortunato, and J{\'a}nos Kert{\'e}sz.
\newblock Detecting the overlapping and hierarchical community structure in
  complex networks.
\newblock {\em New Journal of Physics}, 11(3):033015, 2009.

\bibitem{mucha2010community}
Peter~J Mucha, Thomas Richardson, Kevin Macon, Mason~A Porter, and Jukka-Pekka
  Onnela.
\newblock Community structure in time-dependent, multiscale, and multiplex
  networks.
\newblock {\em Science}, 328(5980):876--878, 2010.

\bibitem{Gaogaodong2018resilience}
Gaogao Dong et~al.
\newblock Resilience of networks with community structure behaves as if under
  an external field.
\newblock {\em Proceedings of the National Academy of Sciences}, page
  201801588, 2018.

\bibitem{(1-6)-huang2009introduction}
Kerson Huang.
\newblock {\em Introduction to statistical physics}.
\newblock Chapman and Hall/CRC, 2009.

\bibitem{(1-4-6)-stanley1971phase}
H~Eugene Stanley.
\newblock {\em Phase transitions and critical phenomena}.
\newblock Clarendon Press, Oxford, 1971.

\bibitem{(4-5)-stauffer2014introduction}
D.~Stauffer and A.~Aharony.
\newblock {\em Introduction to percolation theory: revised second edition}.
\newblock CRC press, 2014.

\bibitem{lee2016hybrid}
Deokjae Lee, S~Choi, M~Stippinger, J~Kert{\'e}sz, and B~Kahng.
\newblock Hybrid phase transition into an absorbing state: Percolation and
  avalanches.
\newblock {\em Physical Review E}, 93(4):042109, 2016.

\bibitem{newman2001random}
Mark~EJ Newman, Steven~H Strogatz, and Duncan~J Watts.
\newblock Random graphs with arbitrary degree distributions and their
  applications.
\newblock {\em Physical review E}, 64(2):026118, 2001.

\bibitem{crisanti2003violation_FDT}
A~Crisanti and F~Ritort.
\newblock Violation of the fluctuation--dissipation theorem in glassy systems:
  basic notions and the numerical evidence.
\newblock {\em Journal of Physics A: Mathematical and General}, 36(21):R181,
  2003.

\bibitem{grigera1999observation_VFDT}
Tom{\'a}s~S Grigera and NE~Israeloff.
\newblock Observation of fluctuation-dissipation-theorem violations in a
  structural glass.
\newblock {\em Physical Review Letters}, 83(24):5038, 1999.

\bibitem{marinari1998violation_FDT}
Enzo Marinari, Giorgio Parisi, Federico Ricci-Tersenghi, and Juan~J
  Ruiz-Lorenzo.
\newblock Violation of the fluctuation-dissipation theorem in
  finite-dimensional spin glasses.
\newblock {\em Journal of Physics A: Mathematical and General}, 31(11):2611,
  1998.

\bibitem{baez2003fluctuation_FDT_metastable}
G~B{\'a}ez, H~Larralde, F~Leyvraz, and RA~M{\'e}ndez-S{\'a}nchez.
\newblock Fluctuation-dissipation theorem for metastable systems.
\newblock {\em Physical Review Letters}, 90(13):135701, 2003.

\bibitem{buldyrev2010catastrophic}
Sergey~V Buldyrev et~al.
\newblock Catastrophic cascade of failures in interdependent networks.
\newblock {\em Nature}, 464(7291):1025, 2010.

\bibitem{dorogovtsev2006k}
Sergey~N Dorogovtsev, Alexander~V Goltsev, and Jose Ferreira~F Mendes.
\newblock K-core organization of complex networks.
\newblock {\em Physical Review Letters}, 96(4):040601, 2006.

\bibitem{goltsev2006k}
Alexander~V Goltsev, Sergey~N Dorogovtsev, and Jose Ferreira~F Mendes.
\newblock k-core (bootstrap) percolation on complex networks: Critical
  phenomena and nonlocal effects.
\newblock {\em Physical Review E}, 73(5):056101, 2006.

\bibitem{carmi2007model}
Shai Carmi, Shlomo Havlin, Scott Kirkpatrick, Yuval Shavitt, and Eran Shir.
\newblock A model of internet topology using k-shell decomposition.
\newblock {\em Proceedings of the National Academy of Sciences},
  104(27):11150--11154, 2007.

\bibitem{schwarz2006onset_kcore_field}
JM~Schwarz, Andrea~J Liu, and LQ~Chayes.
\newblock The onset of jamming as the sudden emergence of an infinite k-core
  cluster.
\newblock {\em EPL (Europhysics Letters)}, 73(4):560, 2006.

\bibitem{kubo1966fluctuation_FDT}
Rep Kubo.
\newblock The fluctuation-dissipation theorem.
\newblock {\em Reports on progress in physics}, 29(1):255, 1966.

\bibitem{weber1956fluctuation_FDT}
J~Weber.
\newblock Fluctuation dissipation theorem.
\newblock {\em Physical Review}, 101(6):1620, 1956.

\end{thebibliography}
\end{document}